\documentclass[epj,nopacs]{svjour}
\usepackage{graphics}
\usepackage{xcolor}

\begin{document}

\title{\boldmath Study of the process
$e^+e^- \to \eta\eta\gamma$ in the energy range $\sqrt{s} = \mbox{1.17--2.00}$ GeV with the SND detector}
\authorrunning{M.~N.~Achasov et al.}
\titlerunning{Process
$e^+e^- \to \eta\eta\gamma$ in the energy range $\sqrt{s} = \mbox{1.17--2.00}$ GeV}

\author{{\large The SND Collaboration}\\ \\
M.~N.~Achasov\inst{1,2} \and
A.~Yu.~Barnyakov\inst{1,2} \and
K.~I.~Beloborodov\inst{1,2} \and
A.~V.~Berdyugin\inst{1,2} \and
D.~E.~Berkaev\inst{1,2} \and
A.~G.~Bogdanchikov\inst{1}\and
A.~A.~Botov\inst{1}\and
T.~V.~Dimova\inst{1,2} \and
V.~P.~Druzhinin\inst{1,2} \and
A.~N.~Kirpotin\inst{1} \and
L.~V.~Kardapoltsev\inst{1,2} \and
A.~S.~Kasaev\inst{1} \and
A.~G.~Kharlamov\inst{1,2} \and
I.~A.~Koop\inst{1,2} \and
A.~A.~Korol\inst{1,2} \and
D.~P.~Kovrizhin\inst{1} \and
A.~S.~Kupich\inst{1,2}\and
K.~A.~Martin\inst{1}\and
N.~A.~Melnikova\inst{1}\and
N.~Yu.~Muchnoy\inst{1,2} \and
A.~E.~Obrazovsky\inst{1}\and
E.~V.~Pakhtusova\inst{1}\and
K.~V.~Pugachev\inst{1,2} \and
D.~V.~Rabusov\inst{1}\and
Yu.~A.~Rogovsky\inst{1,2} \and
A.~I.~Senchenko\inst{1,2} \and
S.~I.~Serednyakov\inst{1,2} \and
D.~N.~Shatilov\inst{1}\and
Yu.~M.~Shatunov\inst{1,2} \and 
D.~A.~Shtol\inst{1} \and
D.~B.~Shwartz\inst{1,2} \and
Z.~K.~Silagadze\inst{1,2} \and
I.~K.~Surin\inst{1} \and
M.~V.~Timoshenko\inst{1} \and
Yu.~V.~Usov\inst{1} \and
V.~N.~Zhabin\inst{1} \and
V.~V.~Zhulanov\inst{1,2}}

\institute{Budker Institute of Nuclear Physics, SB RAS, Novosibirsk, 630090, Russia \and 
Novosibirsk State University, Novosibirsk, 630090, Russia 
}
\date{}
\abstract{
The process $e^+e^-\to\eta\eta\gamma$ is studied in the center-of-mass energy
range 1.17-2.00 GeV using data with an integrated luminosity of 201 pb$^{-1}$ 
collected by the SND detector at the VEPP-2000 $e^+e^-$ collider. 
The $e^+e^-\to\eta\eta\gamma$ cross section is  measured for the first time.
It is shown that the dominant mechanism of this reaction is the transition
through the $\phi\eta$ intermediate state. Our result on the
$e^+e^-\to\eta\eta\gamma$ cross section is consistent with 
the $e^+e^-\to\phi\eta$ measurement in the $\phi\to K^+ K^-$ mode. The search
for radiative processes contributing to the $e^+e^-\to\eta\eta\gamma$
cross section is performed, and no significant signal is observed.}

\maketitle

\section{Introduction\label{intro}}
This work is devoted to study of the process
\begin{equation}
e^+e^- \to \eta\eta\gamma
\label{epg-eq}
\end{equation}
in the center-of-mass (c.m.) energy range $\sqrt{s} = 1.17$--2.00~GeV at the 
experiment with the SND detector at the VEPP-2000 $e^+e^-$ collider.
Previously, this process was studied only near the $J/\psi$ resonance
by the Crystal Ball~\cite{cb_eeg} and BESIII~\cite{bes_eeg} collaborations.

The dominant contribution to the $e^+e^-\to\eta\eta\gamma$ cross section
in the energy region under study comes from the process
$e^+e^-\to\phi\eta$ with the decay $\phi\to\eta\gamma$. This process
was well studied in the SND~\cite{phieta-snd}, CMD-3~\cite{phieta-cmd},
and BABAR~\cite{phieta-babar1,phieta-babar2} experiments in the decay channel 
$\phi\to K^+K^-$. In this work, the most interesting is the search of other
intermediate states. In addition to the known
$\rho\eta$ and $\omega\eta$ mechanisms, the most probable are 
radiative decays of excited vector mesons into $f_{0}(1500)\gamma$ 
and $f_{2}^{\prime}(1525)\gamma$.
Measuring the widths of these decays is important for understanding 
of the quark structure of excited light vector mesons. In particular, 
there are indications that the excited states of $\rho$ and $\omega$ 
mesons can be a mixture of $q\bar{q}$ and vector hybrid 
states~\cite{kalashnikova-hybrid}. The widths of radiative decays of 
excited vector mesons are sensitive to the hybrid state 
admixture~\cite{kalashnikova}. 

In Ref.~\cite{kalashnikova}, the decay width of 
$\phi(1680)\to f_{2}^{\prime}(1525)\gamma$ was calculated within the 
framework of the quark model. For the $\rho(1700)\to f_0(1500)\gamma$ channel
the additional assumption was used that the states $f_0(1370)$, $f_0(1500)$, 
and $f_0(1710)$ are a mixture of states from light quarks and a glueball. 
It was shown that the width of this decay strongly 
depends on the glueball mass. The total cross section for the
production of resonances $\phi(1680)$ and $\rho(1700)$ in $e^+e^-$ collisions
can be estimated as 11 and 15 nb, respectively. Using results of
Ref.~\cite{kalashnikova} and the branching ratios 
$B(f_{2}^{\prime}(1525) \to \eta\eta)=11.6\pm2.2\%$ and
$B(f_{0}(1500) \to \eta\eta)=6.0\pm0.9\%$~\cite{pdg} we obtain 
the following estimates for the cross sections at the maxima of the resonances
$\rho(1700)$ and $\phi(1680)$:
\begin{eqnarray}
\sigma(e^+e^-\to f_2^{\prime}(1525)\gamma \to \eta\eta\gamma)& = &
1.7\mbox{pb},\\
\sigma(e^+e^-\to f_0(1500)\gamma \to \eta\eta\gamma)& = &
0.4\mbox{--}1.9\mbox{ pb}.
\end{eqnarray}
These cross sections are small compared to the cross section 
for the $\phi\eta$ intermediate state, which is about 35 pb at
$\sqrt{s} = 1.68$ GeV. 

\section{Detector and experiment}

SND is a general-purpose non-magnetic detector~\cite{SND_desc,SND_desc2,SND_desc3,SND_desc4} collecting data
at the VEPP-2000 $e^+e^-$ collider~\cite{vepp2k}. Its main part is a 
three-layer spherical electromagnetic calorimeter consisting of 1630 NaI(Tl) 
crystals. The calorimeter covers a solid angle of 95\% of 4$\pi$. The energy
resolution of the calorimeter for photons is
$\sigma_{E}/E=4.2\%/\sqrt[4]{E({\rm GeV})}$. The angular resolution is about
$1.5^\circ$. Directions of charged particles are measured using a nine-layer 
drift chamber and one-layer proportional chamber in a common gas volume. The
solid angle of the tracking system is 94\% of 4$\pi$. A system of threshold
aerogel Cherenkov counters located between the tracking system and the 
calorimeter is used for charged kaon identification. Outside the calorimeter,
a muon detector consisting of proportional tubes and scintillation counters
is placed.

Monte-Carlo (MC) simulation of the signal and background processes takes
into account radiative corrections~\cite{FadinRad}. The angular distribution 
of hard photon emitted from the initial state is generated according to 
Ref.~\cite{BoneMartine}. Interactions of the particles produced in 
$e^+e^-$ annihilation with the detector materials are simulated using
the GEANT4 software~\cite{geant}. The simulation takes into account variation
of experimental conditions during data taking, in particular dead detector 
channels and beam-induced background. To take into account the effect of 
superimposing the beam background on the $e^+e^-$ annihilation events,
simulation uses special background events recorded during data taking with a 
random trigger. These events are superimposed on simulated events, 
leading to the appearance of additional tracks and photons.

The analysis presented in this work is based on data with an
integrated luminosity of 201 pb$^{-1}$ recorded in 2010-2020. 
These data were collected at 138 energy points in the
energy region $\sqrt{s} = 1.17$--2.00~GeV. Since the cross section of
the process under study is small, for the convenience of analysis,
the data are combined into 6 energy intervals listed in Table~\ref{eegtab}.

The process $e^+e^-\to\eta\eta\gamma$ is studied in the 
five-photon final state. Since there are no charged particles 
in the final state of the process under study, it is viable to use the process 
$e^+e^-\to\gamma\gamma$ for normalization. 
As a result of the
normalization a part of systematic uncertainties associated
with the hardware event selection and beam-induced background is canceled out.
Accuracy of the luminosity measurement using the 
process $e^+e^-\to\gamma\gamma$ is estimated to be 2\%~\cite{epg-snd}.

\section{Event selection}
Selection of $e^+e^-\to\eta\eta\gamma\to5\gamma$ events is performed in two 
stages. At the first stage, we select events with exactly 5 photons with
energies above 20 MeV and no charged tracks. The latter condition is ensured by
requiring that the number of hits in the drift chamber is less than four.
The transverse profile of energy deposition in the calorimeter for 
reconstructed photons must be consistent with the expected distribution 
for electromagnetic shower~\cite{chig}. To suppress cosmic-ray background, 
absence of a signal in the muon system is
required. The following conditions provides an approximate balance of energy
and momentum in an event:
\begin{equation}
E_{\rm EMC} / \sqrt{s} > 0.6,\,  P_{\rm EMC} / \sqrt{s} < 0.3,
\end{equation}
where $E_{\rm EMC}$ is the total energy deposition in the calorimeter, and 
$P_{\rm EMC}$ is the total event momentum calculated using energy depositions 
in calorimeter crystals.

The main background processes are 
\begin{equation}
\label{bkglist1}
e^+e^-\to\omega\pi^0\to\pi^0\pi^0\gamma,
\end{equation}
and
\begin{equation}
\label{bkglist0}
e^+e^-\to\omega\eta\to\eta\pi^0\gamma.
\end{equation}
We also study background from the following reactions with multiphoton
final states:
\begin{eqnarray}
e^+e^- &\to& \omega\pi^0\pi^0\to\pi^0\pi^0\pi^0\gamma,\\ \nonumber
e^+e^- &\to& \omega\eta\pi^0\to\eta\pi^0\pi^0\gamma,\label{bkglist3}
\end{eqnarray}
and from the QED processes
\begin{equation}
\label{bkglist2}
e^+e^- \to  4\gamma, 5\gamma.
\end{equation}
Additional photons in the process $e^+e^- \to  4\gamma$ arise from the splitting
of electromagnetic showers, and the beam-induced background.

To suppress background from the processes listed above,
a kinematic fits are performed to the hypotheses 
$ e^+e^-\to 5\gamma$, $e^+e^-\to\pi^0\pi^0\gamma$, $e^+e^-\to\eta\pi^0\gamma$,
and $e^+e^-\to\eta\eta\gamma$ with the requirements of energy and momentum 
balance in an event. For the hypotheses with final states $\pi^0\pi^0\gamma$,
$\eta\pi^0\gamma$, and $\eta\eta\gamma$, the additional constraints are 
imposed that the invariant masses of photon pairs are equal to the masses of 
the $\pi^0$ and $\eta$ mesons.
As a result of the kinematic fit, the energies and 
angles of photons are refined, and the $\chi^2$ of the 
proposed kinematic hypothesis is calculated. In the kinematic fits,
all possible combinations of photons are tested, and the combination with the
smallest $\chi^2$ value is retained. The following conditions are 
applied to the obtained $\chi^2$ values 
\begin{equation}
\chi^2_{\eta\eta\gamma} - \chi^2_{5\gamma} < 60, \quad
\chi^2_{5\gamma} < 30,
 \quad \chi^2_{\pi^0\pi^0\gamma} - \chi^2_{5\gamma} > 100.
 \label{cuts1}
\end{equation}

For events with $\sqrt{s}>$1.32 GeV, the additional condition is imposed
\begin{equation}
\chi^2_{\eta\pi^0\gamma} - \chi^2_{5\gamma} > 30.
\label{cuts2}
\end{equation}
Using the criteria desctibed above, 183 data events are selected for
further analysis. 

\section{\boldmath Fitting the $\chi^2_{\eta\eta\gamma} - \chi^2_{5\gamma}$ 
distribution}
\begin{figure}
\centering
\resizebox{0.45\textwidth}{!}{\includegraphics{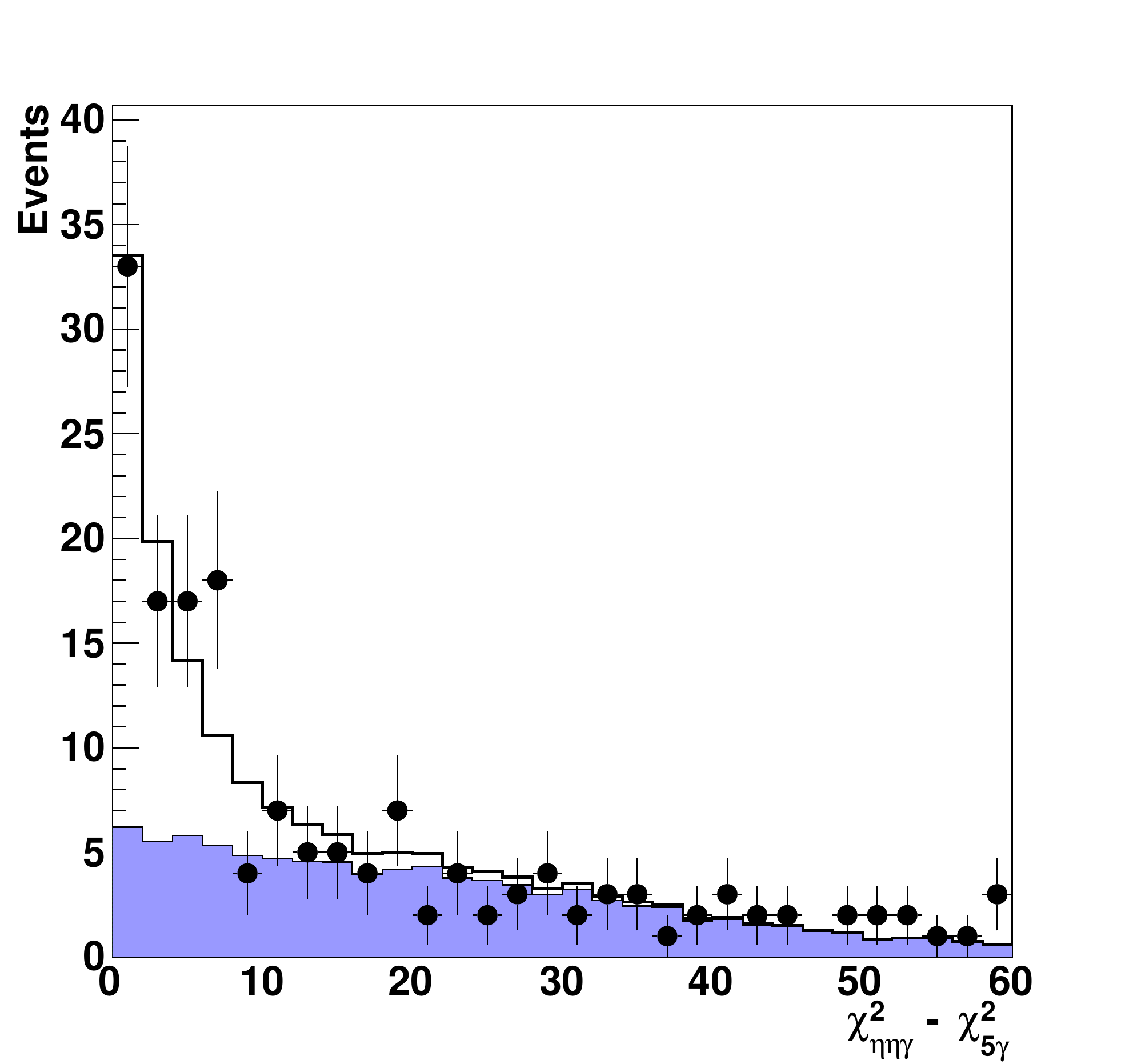}}
\caption{The $\chi^2_{\eta\eta\gamma}-\chi^2_{5\gamma}$ distribution for
selected data events with $\sqrt{s}=1.57$--2.00 GeV (points with error bars). 
The solid histogram is the result of the fit with the sum of the signal and 
background distributions described in the text. The shaded histogram
represents the background distribution.
\label{FitRes}}
\end{figure}
The $\chi^2_{\eta\eta\gamma} - \chi^2_{5\gamma}$ distribution 
for selected data events from the energy range $\sqrt{s}=1.57$--2.00 GeV is
shown in Fig.~\ref{FitRes}. 
To determine the number of $\eta\eta\gamma$ events, this distribution is
fitted with the sum of signal and background distributions. 

The signal distribution is obtained using MC simulation
as a sum of the distributions for the processes 
\begin{equation}
e^+e^-\to\phi\eta, \quad e^+e^-\to\rho\eta,\quad e^+e^-\to\omega\eta
\label{eeg-vp}
\end{equation}
with the subsequent decays of the vector mesons to $\eta\gamma$. 
The interference between these processes is neglected in the sum.
The total number of signal events is a free fit parameter. The fractions of
the processes (\ref{eeg-vp}) in the sum are calculated using their measured
cross sections~\cite{phieta-cmd,rhoeta-babar,ometa-snd} and the branching 
fractions of the decays $\phi$, $\rho$, $\omega\to\eta\gamma$~\cite{pdg}.

Imperfect simulation of the detector response may lead to difference between
data and simulation in the $\chi^2_{\eta\eta\gamma} - \chi^2_{5\gamma}$
distribution. To study this effect, we use events of the process 
$e^+e^-\to\pi^0\pi^0\gamma$ from the energy region $\sqrt{s} = 1.05$--1.7 GeV,
where they are selected practically without background. For these events, the
distribution of the parameter $\chi^2_{\pi^0\pi^0\gamma} - \chi^2_{5\gamma}$
is studied.  The parameters $\chi^2_{\eta\eta\gamma}$ and 
$\chi^2_{\pi^0\pi^0\gamma}$ are calculated using the kinematic fits with the 
same number of constraints, differing only in the masses of intermediate
particles. Although the photon spectra in  $\eta\eta\gamma$ and 
$\pi^0\pi^0\gamma$ events are different, we expect that the correction to 
the $\chi^2_{\pi^0\pi^0\gamma} - \chi^2_{5\gamma}$ distribution
is close to the correction to the $\chi^2_{\eta\eta\gamma} - \chi^2_{5\gamma}$
distribution. It turned out that the difference between data and simulation
can be described by stretching the simulated distribution.
To find the stretching parameter $\alpha_b$, the data distribution is fitted
by the simulated distribution of the parameter
$\alpha_b(\chi^2_{\pi^0\pi^0\gamma} - \chi^2_{5\gamma})$.
The resulting value $\alpha_b = 1.05\pm0.01$ allows to match well the 
distributions for data and simulation. In particular,
$\chi^2/{\rm ndf}$ changes from 49/29 at $\alpha_b=1$ to 34/28 at 
$\alpha_b=1.05$, where ${\rm ndf}$ is the number of degrees of freedom.
This correction is applied to the $\chi^2_{\eta\eta\gamma} - \chi^2_{5\gamma}$ 
distribution obtained from simulation. To estimate the associated
systematic uncertainty, $\alpha_b$ is treated in the fit as a nuisance 
parameter with a Gaussian distribution, the standard deviation of which is conservatively
estimated to be 0.05. 

For description of the background distribution, the processes
(\ref{bkglist1}-\ref{bkglist2}) are taken into account.
Other multiphoton processes make a negligible contribution. 
The total number of background events is a free fit parameter, while the 
relative contributions ($\beta_i$) of the different processes are calculated
using their measured cross sections. Since the 
$\chi^2_{\eta\eta\gamma} - \chi^2_{5\gamma}$ distributions 
for different background processes ($P_{\rm bkg, i}$) are slightly different,
the uncertainties in $\beta_i$ lead to an uncertainty in background 
subtraction. To estimate this uncertainty, the background distribution used in
the fit is presented as a sum $\sum_i \beta_i P_{\rm bkg, i}$. The
parameters $\beta_i$ are treated as nuisance parameters with Gaussian 
constraints.

The result of the fit to the $\chi^2_{\eta\eta\gamma} -\chi^2_{5\gamma}$ 
distribution for data events from the energy range $\sqrt{s}=1.57$--2.00 GeV
is shown in Fig.~\ref{FitRes}.
The following values are obtained for the numbers of the signal and
background events 
\begin{equation}
N_{\eta\eta\gamma} = 69.7 \pm 12.0, \quad N_{\rm bkg} = 91.3 \pm 12.9.
\label{NevtFit}
\end{equation}
The quoted errors are statistical. The systematic uncertainty in the number 
of signal events due to the uncertainty in $\alpha_b$ is 3\%.
The uncertainty due to the variation of the $\beta_i$ coefficients is 1\%.
The total contribution from background 
processes~(\ref{bkglist1}-\ref{bkglist3}) calculated using their measured
cross sections and the detection efficiencies obtained from simulation
is $N_{\rm bkg}^{\rm exp} = 81.4 \pm 4.3$.
This value is in good agreement with the result of the fit. 

The fit described above is performed in 6 energy intervals. 
The results are listed in Table~\ref{eegtab}.
\begin{table*}
\center
\caption{
The energy range ($\sqrt{s}$), integrated  luminosity  ($L$), detection 
efficiency ($\varepsilon$), number of signal ($N_{\eta\eta\gamma}$) and 
background ($N_{\rm bkg}$) events obtained from the fit, number of expected 
background events ($N^{\rm exp}_{\rm bkg}$), radiative correction ($1+\delta$) and
Born cross section ($\sigma$) for the process  $e^+e^-\to  \eta\eta\gamma$.
The quoted errors are statistical. The systematic uncertainty in the cross
section is 21\% for $\sqrt{s}< 1.32$ GeV, 23\% for $1.32<\sqrt{s}<1.57$ GeV,
and 12\% for $\sqrt{s}>1.57$ GeV. 
\label{eegtab}}
\begin{tabular}{|c|c|c|c|c|c|c|c|}
\hline
$\sqrt{s}$, GeV & $L$, pb$^{-1}$ & $\varepsilon$, \% & $N_{\eta\eta\gamma}$ &
$N_{\rm bkg}$ &$N^{\rm exp}_{\rm bkg}$ & $1+\delta$ & $\sigma$,\\  \hline 
1.17-1.32 &  26.0 & 4.81 & 1.4  $\pm$ 3.4 & 17.6 $\pm$ 5.2 & 16.9 $\pm$ 0.9 & 0.754 & 1.4 $\pm$ 3.6\\ 
1.32-1.57 &  27.4 & 3.10 & -1.4 $\pm$ 1.6 & 3.9  $\pm$ 1.8  & 9.6 $\pm$ 0.4 & 0.881 & -1.9 $\pm$ 2.1\\ 
1.57-1.70 &  19.3 & 4.06 & 18.8 $\pm$ 5.3 & 13.2 $\pm$ 4.7 & 12.0 $\pm$ 0.7 & 0.857 & 28.0 $\pm$ 7.9\\ 
1.70-1.80 &  15.8 & 4.18 & 14.8 $\pm$ 4.5 &  6.2 $\pm$ 3.4 &  8.6 $\pm$ 0.7 & 0.972 & 23.0 $\pm$ 7.0\\ 
1.80-1.90 &  65.6 & 3.84 & 14.6 $\pm$ 6.2 & 39.4 $\pm$ 7.9 & 35.9 $\pm$ 4.7 & 1.062 & 5.4 $\pm$ 2.3\\ 
1.90-2.00 &  46.8 & 3.83 & 20.9 $\pm$ 7.4 & 33.1 $\pm$ 8.2 & 25.0 $\pm$ 4.5 & 1.095 & 10.6 $\pm$ 3.8\\ 
\hline
\end{tabular}
\end{table*}

\section{Detection efficiency and radiative corrections}
\begin{figure}
\centering
\resizebox{0.48\textwidth}{!}{\includegraphics{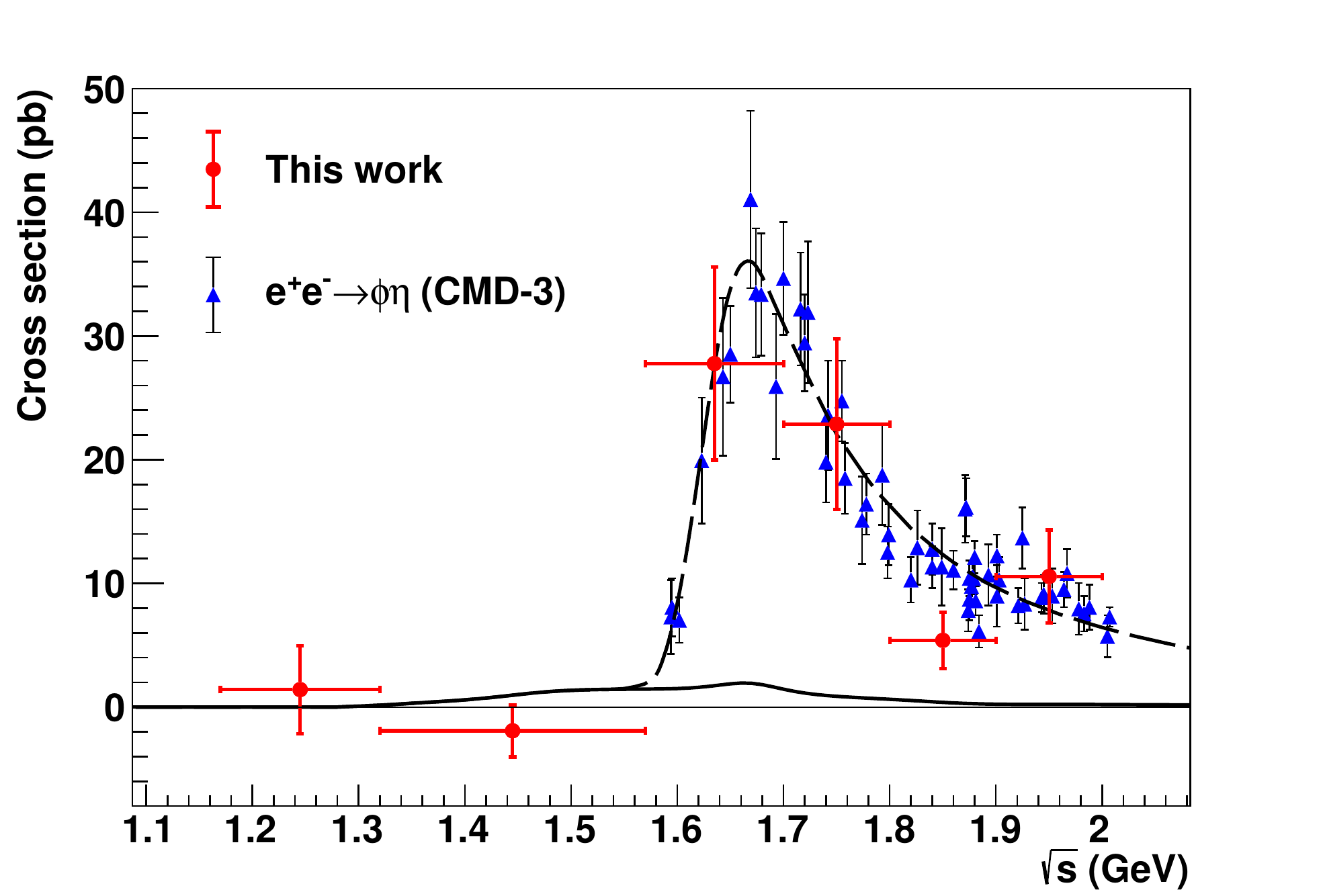}}
\caption{The the $e^+e^-\to \eta\eta\gamma$ Born cross section measured in
this work (circles) compared with the $e^+e^-\to \phi\eta$ cross section
measured by CMD-3 in the decay mode $\phi\to K^+K^-$~\cite{phieta-cmd}
multiplied by the branching fraction of the decay $\phi\to\eta\gamma$
(triangles). The solid curve is the sum of the $e^+e^-\to \rho\eta$ and
$\omega\eta$ cross sections, while the dashed line is the sum of the
$e^+e^-\to \phi\eta$, $\rho\eta$ and $\omega\eta$ cross sections.
\label{CrsPic}}
\end{figure}

The Born cross section averaged over the energy interval is calculated as 
follows 
\begin{equation}
\sigma = \frac{N_{\eta\eta\gamma}}{ L \varepsilon (1+\delta)},
\label{avgcrs}
\end{equation}
where $L=\sum_i L_i$ is the total integrated luminosity for the interval,
$L_i$ is the integrated luminosity at the $i$th energy point within the 
interval, $\varepsilon$ and $\delta$ are the detection efficiency and 
radiative correction averaged over the interval. 
To calculate $\varepsilon$ and $\delta$, it is necessary to know 
relations between different intermediate mechanisms of the process 
$e^+e^-\to \eta\eta\gamma$ and the energy dependence of the cross section 
for each of the mechanisms. We assume that the intermediate mechanisms 
$\phi\eta$, $\omega\eta$ and $\rho\eta$ make the dominant contribution 
to the cross section of the process under study. The Born cross section 
calculated as the sum of the contributions of these mechanisms is shown 
in Fig.~\ref{CrsPic}. The calculation is based on the approximation of the
experimental data~\cite{phieta-cmd,rhoeta-babar,ometa-snd}.

Above 1.6 GeV, the main contribution to the cross section arises from the 
$\phi\eta$ mechanism, while below 1.6 GeV, the dominant mechanism is 
$\rho\eta$. In the sum, we neglect the interference between the amplitudes of 
the intermediate states. The highest interference is expected between
amplitudes $\rho\eta$ and $\omega\eta$. Its value depends on the relative 
phase between the amplitudes. We estimate that at 1.5 GeV the maximum value 
of the interference contribution to the cross section can reach 30\%. 
At 1.65 GeV, the interference contribution does not exceed 1\%.

For each energy point $i$ and each reaction mechanism $j$ we calculate the 
visible cross section
\begin{equation}
\sigma_{vis,j}(s_i) = \int\limits_{0}^{x_{max}}F(x,s_i)
\sigma_j(s_i (1-x))dx,
 \label{viscrs}
\end{equation}
where $F(x,E)$ is a so-called radiator function~\cite{FadinRad} describing the
distribution of the energy fraction $x=E_r/2\sqrt{s}$ carried away by photons
emitted from the initial state, and $\sigma_j(s)$ is the Born cross section
for $j$th mechanism.

The detection efficiency and radiative correction averaged over the 
energy interval are calculated as
\begin{eqnarray}
\varepsilon &=& \sum \limits_{i,j} \varepsilon_{i,j} L_i \sigma_{vis,j}(s_i)
\Big/
\sum \limits_{i,j} L_i \sigma_{vis,j}(s_i), \label{raddef1}\\
1+\delta &=& \sum \limits_{i,j} L_i \sigma_{vis,j}(s_i)\Big/
\sum \limits_{i,j} L_i \sigma_j(s_i),
\label{raddef2}
\end{eqnarray}
where $\varepsilon_{i,j}$ is the detection efficiency calculated using 
simulation for the $j$th intermediate mechanism at the $i$th energy point. 
The obtained values of $\varepsilon$ and $(1+\delta)$ are listed in 
Table~\ref{eegtab}.

To estimate the model dependence of the measured cross section, the 
$\phi\eta$, $\omega\eta$  and $\rho\eta$ contributions in Eqs.~(\ref{raddef1})
and (\ref{raddef2}) are varied within their experimental errors.
To determine the uncertainty associated with the interference between the 
$\omega\eta$ and $\rho\eta$ mechanisms, we increased/decreased the 
$\omega\eta$ contribution by the value of the interference term between the 
$\omega\eta$ and $\rho\eta$ amplitudes.
The efficiency is weakly dependent on the intermediate mechanism. 
Therefore, when calculating the efficiency, we assume that the efficiency for
events corresponding to the interference term is close to the efficiency for 
$\omega\eta$ events. We also added to the sums (\ref{raddef1}) and 
(\ref{raddef2}) terms corresponding to the
mechanisms $f_{0}(1500)\gamma$ or $f_{2}^{\prime}(1525)\gamma$ with 
cross sections equal to the upper limits set in Sec.~\ref{raddec}. The model
uncertainty in the $e^+e^-\to \eta\eta\gamma$ cross section determined in this
way does not exceed 2\% for $\sqrt{s}> 1.57$ GeV and 20\% for $\sqrt{s}<1.57 $
GeV. 

Imperfect simulation of the detector response leads to a systematic 
uncertainty in the detection efficiency obtained using the simulation.
The difference between data and simulation in the distributions of $\chi^2$ of 
kinematic fits for the process $e^+e^-\to\omega\pi^0\to\pi^0\pi^0\gamma$ were
studied using large statistics in Ref.~\cite{ompi,ompi2}. Based on this study, we
conclude that the systematic uncertainty due to the conditions~(\ref{cuts1})
does not exceed 5\%. 
The condition $\chi^2_{\eta\pi\gamma}-\chi^2_{5\gamma}> 30 $,
specific for this analysis, reduces the detection efficiency by approximately
2 times.
To determine the efficiency correction associated 
with this condition, its boundary is varied from 30 to 10. The observed
variation of the measured cross section is $1.01\pm 0.10$. 
Thus, no additional efficiency correction is introduced, 
and the error of the obtained correction 10\% is taken as an estimate of 
the systematic uncertainty due to condition~(\ref{cuts2}). The total 
systematic uncertainty of the detection efficiency 
due to the selection conditions is 11\% for $\sqrt{s}> 1.32$ GeV and 
5\% for $\sqrt{s}< 1.32$ GeV. 

In SND, photons converting into a $e^+e^-$ pair in the material
before the drift chamber, produce a charged track. Such events do not pass 
the selection criteria.
Since the process under study and the process used for normalization contain 
different numbers of photons in the final state, improper simulation of 
the photon conversion lead to a shift in the measured cross section. 
The photon conversion probability is measured using events of the 
process $e^+e^-\to\gamma\gamma$. The corresponding efficiency correction
is found to be $(-0.79 \pm 0.02)\%$.

The values of the $e^+e^-\to\eta\eta\gamma$ Born cross section
for six energy intervals obtained using Eq.~(\ref{avgcrs}) are listed in 
Table~\ref{eegtab}. In Fig.~\ref{CrsPic}, the measured cross section is 
compared with the calculation for the mechanisms $\phi\eta$, $\omega\eta$ and
$\rho\eta$ based on the approximation of the experimental 
data~\cite{phieta-cmd,rhoeta-babar,ometa-snd}. Figure~\ref{CrsPic} also shows 
the measurement of the cross section for the dominant mechanism 
$e^+e ^-\to\phi\eta$ performed by CMD-3 in decay mode 
$\phi\to K ^+K^-$~\cite{phieta-cmd}. It is seen that our results agree with 
both the calculation and the CMD-3 measurement. 

Only statistical errors are quoted in Table~\ref{eegtab} and Fig.~\ref{CrsPic}.
The systematic uncertainty is 21\% for $\sqrt{s}< 1.32$ GeV, 23\% for 
$1.32<\sqrt{s}<1.57$ GeV, and 12\% for $\sqrt{s}> 1.57$ GeV.
The sources of the systematic uncertainty are listed in Table~\ref{CrsSyst}. 
\begin{table}
\center
\caption{The main sources of the systematic uncertainty in the measured 
$e^+e^-\to \eta\eta\gamma$ cross section.
\label{CrsSyst}}
\begin{tabular}{|l|c|}
\hline
Source&   \\  \hline 
Luminosity & 2\%  \\
Selection conditions & 5-11\% \\
Background subtraction & 1\% \\
$\chi^2_{\eta\eta\gamma} - \chi^2_{5\gamma}$ distribution shape& 3\% \\
Model dependence & 2-20\% \\
Total & 12-23\%  \\
\hline
\end{tabular}
\end{table}

\section{Search for radiative processes \label{raddec}}
\begin{figure}
\centering
\resizebox{0.45\textwidth}{!}{\includegraphics{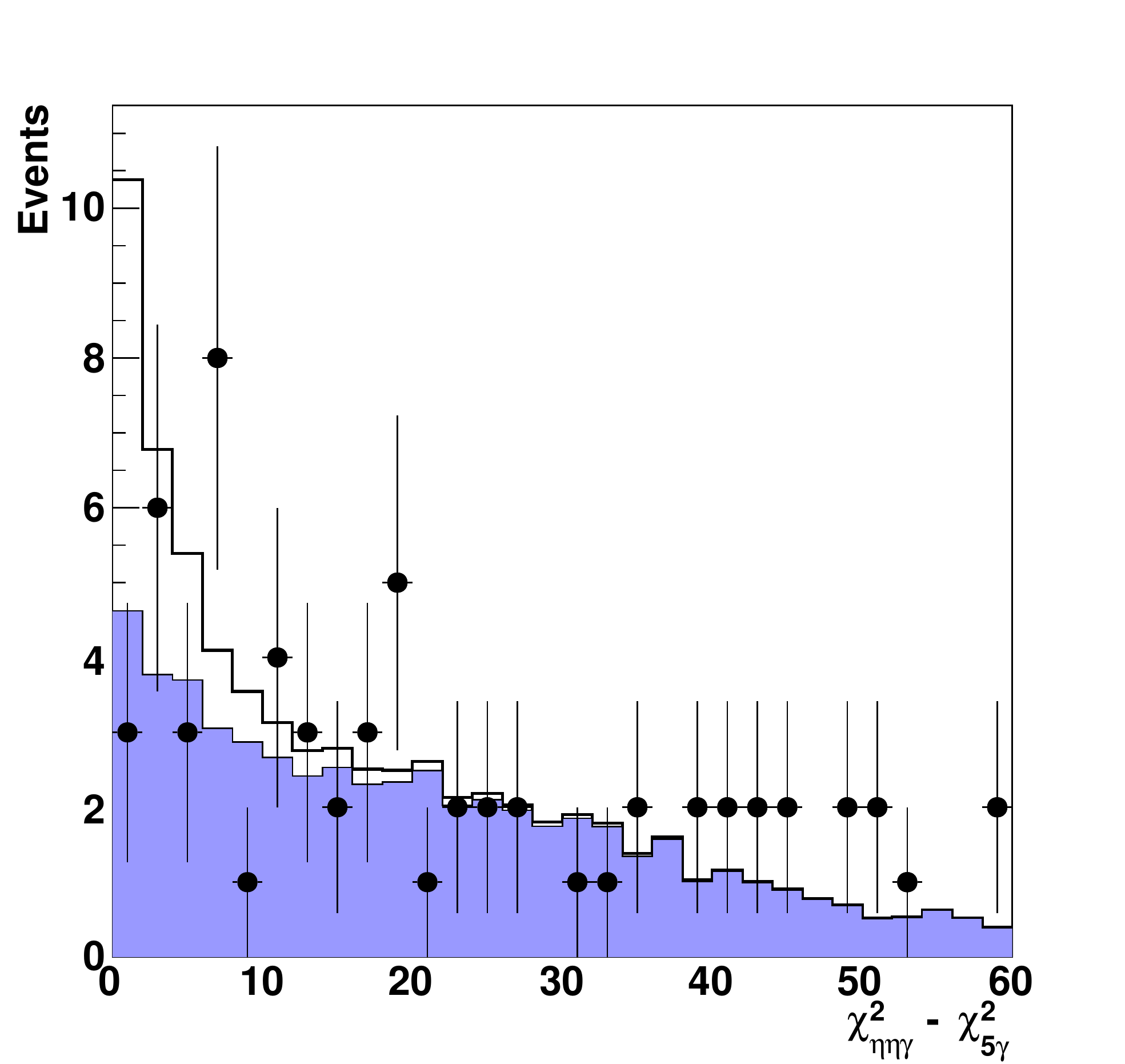}}
\caption{The $\chi^2_{\eta\eta\gamma}-\chi^2_{5\gamma}$ distributions for data
events from the energy region $\sqrt{s} = 1.57$ - 2.00 GeV selected with the
additional condition $\chi^2_{\phi \eta} - \chi^2_{\eta\eta\gamma} > 20$ 
(points with error bars). The shaded histogram is the expected distribution
for the background processes. The solid 
histogram is the sum of the background distribution and the distribution for
events of the process $e^+e^-\to f_{0}(1500)\gamma \to\eta\eta\gamma$ with 
a cross section of 5 pb.
\label{ResNM}}
\end{figure}
It is interesting to search for intermediate mechanisms of the
$e^+e^-\to\eta\eta\gamma$ reaction other than $\phi\eta $, $\omega\eta$, 
and $\rho\eta$. The most likely additional mechanisms are
$f_{0}(1500)\gamma$ and $f_{2}^{\prime}(1525)\gamma$.
To increase the sensitivity to them, it is required to suppress 
the contribution of the the dominant mechanism $\phi\eta$. For this, 
an additional kinematic fit is performed to the hypothesis 
$e^+e^- \to \phi \eta \to \eta\eta\gamma$, in which
the constraint is added that one of the two invariant masses of the 
$\eta\gamma$ system is equal to the $\phi$ meson mass.
The following condition is imposed on the $\chi^2$ of the kinematic fit:
\begin{equation}
\chi^2_{\phi \eta} - \chi^2_{\eta\eta\gamma} > 20.
\label{cuts3}
\end{equation}
This condition rejects about 90\% of $e^+e^-\to\phi\eta$ events.
At the same time, the detection efficiency for the processes 
$e^+e^- \to f_{0}(1500)\gamma$ and $f_{2}^{\prime}(1525)\gamma$
decreases only by a factor of about 2. The number of selected events in 
the entire energy range decreases to 86 after applying this condition. 
The $\chi^2_{\eta\eta\gamma}-\chi^2_{5\gamma}$ distribution for data events 
from the energy region $\sqrt{s} = 1.57$ - 2.00 GeV is shown in 
Fig.~\ref{ResNM}. It is seen that the data distribution is well described 
by the distribution for expected background. 

Since no significant signal from radiative processes is observed, we set 
upper limits on their cross sections. 
To do this, we compare the data $\chi^2_{\eta\eta\gamma}-\chi^2_{5\gamma}$ 
distribution with the same distributions for signal and background,
using the $CL_{s}$ technique~\cite{CLs_desc,CLs_desc2}.

As input, the procedure~\cite{CLs_desc2} requires the measured distribution, 
the distributions for signal and background, the expected number of 
background events, and the systematic uncertainties for the signal and
background. The systematic uncertainties due to the luminosity measurement, 
selection criteria, and imperfect simulation of the 
$\chi^2_{\eta\eta\gamma}-\chi^2_{5\gamma}$ distribution
are similar to those listed in Table~\ref{CrsSyst}.
To estimate the expected number of background events, we use the measured
cross sections for the background processes and the detection efficiencies
obtained using simulation. In this case, background processes are not only 
the processes (\ref{bkglist1}-\ref{bkglist2}), but also (\ref{eeg-vp}). 
\begin{table*}[th!]
\center
\caption{
The energy range ($\sqrt{s}$), integrated  luminosity  ($L$), detection
efficiency ( $\varepsilon$), 90\% CL upper limit on number of events 
of radiative processes ($N_{rad}$),
number of expected background events ($N^{\rm exp}_{\rm bkg}$),
radiative correction ($1+\delta$), 90\% CL upper limit on the Born cross 
section for the radiative processes ($\sigma$).
\label{radtab}}
\begin{tabular}{|c|c|c|c|c|c|c|}
\hline
$\sqrt{s}$, GeV & $L$, pb$^{-1}$ & $\varepsilon$, \% & $N_{rad}$  &
$N^{\rm exp}_{\rm bkg}$ 
& $1+\delta$ & $\sigma$, pb  \\  \hline 
1.17-1.32 &  26.0 & 4.28 & 7.0  & 16.9 $\pm$ 0.9 & 0.915 & 6.8\\ 
1.32-1.57 &  27.4 & 2.60 & 2.5 & 10.3 $\pm$ 0.4 & 0.964 & 3.7\\ 
1.57-1.80 &  35.1 & 2.03 & 7.5 & 17.8 $\pm$ 0.7 & 0.982 & 10.7\\ 
1.80-2.00 &  112.4 & 2.04 & 7.7 & 37.4 $\pm$ 3.5 & 0.989 & 3.4\\ 
\hline
\end{tabular}
\end{table*}

The estimation of the systematic uncertainty in the expected number of 
background events is performed under assumption that it is completely 
determined by the experimental accuracy of the measured cross sections for 
the background processes. The detection efficiency for signal events is
calculated as a half sum of the efficiencies for 
$e^+e^-\to f_{0}(1500)\gamma$ and $f_{2}^{\prime}(1525)\gamma$ events.
Their half difference is used as an estimate of the model dependence 
of the detection efficiency. It does not exceed 6\%. 

To average the efficiency and radiative correction in 
Eqs.~(\ref{raddef1}) and (\ref{raddef2}), we use the assumption that the Born 
cross section for the radiative process does not depend on energy.

The last 4 energy intervals listed in Table~\ref{eegtab} are merged into 2 to
increase statistics.
The obtained values of 90\% confidence level (CL) upper limits for 4 energy
intervals are listed in Table~\ref{radtab}.

\section {Conclusion}
In the experiment with the SND detector at the VEPP-2000 collider the cross
section of the $e^+e^-\to\eta\eta\gamma$ process has been measured in 
the c.m. energy range from 1.17 to 2.00 GeV. The main intermediate mechanism 
in this energy range is $\phi\eta$. The measured $e^+e^-\to\eta\eta\gamma$ 
cross section is in good agreement with the CMD-3 measurement of the
$e^+e^-\to \phi\eta$ process made in the $\phi\to K^+K^-$ decay channel.
A search for contributions to the cross section from radiative processes 
has been carried out. No significant signal has been found. In the energy 
region corresponding to the resonances $\phi(1680)$ and $\rho(1700)$, the 
upper limit is 10.6 pb and significantly exceeds the estimates of the cross
sections based on the predictions of Ref.~\cite{kalashnikova}
(see Sec.~\ref{intro}),
$\sigma(e^+e^-\to\phi(1680)\to f_{2}^{\prime}(1525)\gamma\to \eta\eta\gamma)
=1.7$ pb and
$\sigma(e^+e^-\to\rho(1700)\to f_0(1500)\gamma \to \eta\eta\gamma)
= 0.4$-1.9 pb.


\begin{thebibliography}{99}

\bibitem{cb_eeg}
C.~Edwards {\it et al.} (Crystal Ball Collaboration),
Phys. Rev. Lett. \textbf{48}, 458 (1982)

\bibitem{bes_eeg}
M.~Ablikim \textit{et al.} (BESIII Collaboration),
Phys. Rev. D \textbf{87}, 092009 (2013)
[erratum: Phys. Rev. D \textbf{87}, 119901 (2013)]

\bibitem{phieta-snd}
M.~N.~Achasov \textit{et al.} (SND Collaboration),
Phys. Atom. Nucl. \textbf{81}, 205 (2018)

\bibitem{phieta-cmd}
V.~L.~Ivanov \textit{et al.} (CMD-3 Collaboration),
Phys. Lett. B \textbf{798}, 134946 (2019)

\bibitem{phieta-babar1}
B.~Aubert \textit{et al.} (BaBar Collaboration), 
Phys. Rev. D \textbf{76}, 092005 (2007)
[erratum: Phys. Rev. D \textbf{77}, 119902 (2008)]

\bibitem{phieta-babar2}
B.~Aubert \textit{et al.} (BaBar Collaboration),
Phys. Rev. D \textbf{77}, 092002 (2008)


\bibitem{kalashnikova-hybrid} 
A.~Donnachie and Y.~S.~Kalashnikova,
Phys.\ Rev.\ D {\bf 60}, 114011 (1999)

\bibitem{kalashnikova} 
F.~E.~Close, A.~Donnachie and Y.~S.~Kalashnikova,
Phys.\ Rev.\ D {\bf 65}, 092003 (2002)

\bibitem{pdg}
P.~A.~Zyla \textit{et al.} (Particle Data Group), Review of Particle Physics,
PTEP \textbf{2020},  083C01 (2020)

\bibitem{SND_desc}
M.~N.~Achasov {\it et al.,}, Nucl. Instrum. Meth. A {\bf 598}, 31 (2009)
\bibitem{SND_desc2}
V.~M.~Aulchenko {\it et al.,} Instrum. Meth. A {\bf 598}, 102 (2009)
\bibitem{SND_desc3}
A.~Yu.~Barnyakov {\it et al.,} Nucl. Instrum. Meth. A {\bf 598}, 163 (2009)
\bibitem{SND_desc4}
V.~M.~Aulchenko {\it et al.,}  Nucl. Instrum. Meth. A {\bf 598}, 340 (2009)

\bibitem{vepp2k}
P.~Y.~Shatunov {\it et al.},
Status and perspectives of the VEPP-2000,
Phys.\ Part.\ Nucl.\ Lett.\  {\bf 13}, 995 (2016)

\bibitem{FadinRad}
E.~A.~Kuraev and V.~S.~Fadin,
Sov.\ J.\ Nucl.\ Phys.\  {\bf 41}, 466 (1985)
[Yad.\ Fiz.\  {\bf 41}, 733 (1985)]

\bibitem{BoneMartine}G.~Bonneau and F.~Martin, 
Nucl. Phys. B {\bf 27}, 381 (1971)

\bibitem{geant}
S.~Agostinelli {\it et al.} (GEANT4 Collaboration),
Nucl.\ Instrum.\ Meth.\ A {\bf 506}, 250 (2003)


\bibitem{epg-snd}
M.~N.~Achasov \textit{et al.} (SND Collaboration),
Eur. Phys. J. C \textbf{80},  1008 (2020)

\bibitem{chig}
A.~V.~Bozhenok, V.~N.~Ivanchenko and Z.~K.~Silagadze, 
Nucl. Instrum. Meth. A \textbf{379}, 507 (1996)

\bibitem{rhoeta-babar}
J.~P.~Lees \textit{et al.} (BaBar Collaboration),
Phys. Rev. D \textbf{97}, 052007 (2018)

\bibitem{ometa-snd}
M.~N.~Achasov \textit{et al.} (SND Collaboration),
Phys. Rev. D \textbf{99}, 112004 (2019)

\bibitem{ompi}
M.~N.~Achasov \textit{et al.} (SND Collaboration),
Phys. Rev. D \textbf{88}, 054013 (2013)
\bibitem{ompi2}
M.~N.~Achasov \textit{et al.} (SND Collaboration),
Phys. Rev. D \textbf{94}, 112001 (2016)

\bibitem{CLs_desc}
A.~L.~Read, J. Phys. G \textbf{28}, 2693 (2002)

\bibitem{CLs_desc2}
T.~Junk, Nucl. Instrum. Meth. A \textbf{434}, 435 (1999)

\end{thebibliography}
\end{document}